\documentclass[12pt]{revtex4}
\usepackage{bm}
\usepackage{amsmath}
\usepackage{amssymb}
\usepackage{graphicx}
\usepackage{epsfig}
\begin{document}
\author{U.Sandler \\
Lev Academic Center (JCT), Jerusalem 91160, Israel}

\title{Evolutionary quantization of matter and
Universe expansion.}

\begin{abstract}
In this paper we consider generalization of classical/quantum mechanics that directly follows from the causality principle and topology of a system's state space. In generalized mechanics, the Hamiltonian/Schrodinger equations remain the same, but the Hamiltonian may depend on the action and its canonically conjugated variables as additional dynamical variables. This extension of quantum mechanics indicates that the quantization of matter could be an evolutionary process, and in the distant future, even massive bodies may become entirely quantum objects without well-defined trajectories and shapes. In the classical limit, the first approximation of the Hamiltonian with respect to the action explains the accelerated expansion of the Universe, Hubble’s  law, formation of spiral galaxies with a non-Kepler curve of rotation velocity, and asymmetry between distributions of matter and antimatter. This theory predicts that our open universe could have extended pre-history and be preceded by a long set of closed precursor universes.

\end{abstract}
\maketitle

It was shown recently \cite{s-langr},\cite{sndlr-book} that there are two extension of the Hamiltonian dynamics, which directly follows from the causality principal and common topology of a system's state space (see Appendix \ref{ApSetUp} for details). First of them is equivalent to the statistical mechanics \cite{sndlr-tsitol}, while the second one was not be considered early. In this generalization, the Hamiltonian equations remain the same, but extended Hamiltonian may depend on the action and its canonically conjugated variable (\textit{AC-variable} for brief) as additional dynamical variables. Alternatively, this generalisation of the Hamiltonian dynamics can be obtained by using principal of least action for the "Supper Action" - $\mathbb{S}$ (\textit{S-action} for brief):
\begin{equation}\label{E0.1}
  \mathbb{S}=\mathbb{S}_0+\int_0^t\left[(\bm{\pi}\cdot\bm{\dot{x}})
  +S\dot{w}-\mathcal{H}(\bm{\pi},\bm{x},w,S,t)\right]dt
\end{equation}
where we have taken into account that S-Hamiltonian could depend on the common action - $S$ and its canonically conjugated AC-variable - $w$. We have designated $\bm{x}$ and $t$ as the system's coordinated and time. $\bm{\pi}$ is a "supper momentum" (\textit{S-momentums} for brief), which is related with the common momentum $\bm{p}$ as $\bm{\pi}=w\bm{p}$ and
$\mathcal{H}$ is "super Hamiltonian" (\textit{S-Hamiltonian} for brief), which is related with the common Hamiltonian
$H\left(\bm{p},\bm{x},t\right)$ as
\begin{equation}\label{E0.1a}
 \mathcal{H}=w\left[H\left(\frac{\bm{\pi}}{w},\bm{x},t\right)+
 \mbox{S-dependent terms}\right].
\end{equation}

Variation of (\ref{E0.1}) with respect to all variables leads to the Hamiltonian equations of motion
\begin{eqnarray} \nonumber
  \frac{d\bm{x}}{dt} &=& \frac{\partial\mathcal{H}}
  {\partial \bm{\pi}}, \\  \nonumber
  \frac{d\bm{\pi}}{dt} &=& -\frac{\partial\mathcal{H}}{\partial \bm{x}},
  \\  \nonumber
  \frac{dw}{dt} &=& \frac{\partial\mathcal{H}}{\partial S},
  \\ \nonumber
  \frac{dS}{dt} &=& -\frac{\partial\mathcal{H}}{\partial w}\;\;=\;\;L.
\end{eqnarray}
(where $L$ is the Lagrangian) and to relations between S-action and the common action $S$, AC-variable $w$ and the S-momentum $\bm{\pi}$ (see Appendix \ref{ApSetUp})
\begin{eqnarray}\label{E0.2}
            S &=& \frac{\partial \mathbb{S}}{\partial w}; \;\;\;\;
     \bm{\pi} \;\;=\;\; \frac{\partial \mathbb{S}}{\partial\bm{x}};\;\;\;\;
  \mathcal{H} \;\;=\;\; - \frac{\partial \mathbb{S}}{\partial t}
\end{eqnarray}

\section{How massive body could become a quantum object}\label{QM}

Canonical quantization of the S-Hamiltonian dynamics leads to the commutative relations
\begin{eqnarray}\label{E1.1}
  [\bm{\hat{x}},\bm{\hat{\pi}}] &=& i\hbar; \;\;\;\;
  [\bm{\hat{x}},\bm{\hat{p}}]\;\;=\;\; i\hbar \hat{w}^{-1};\;\;\;\;
  [\hat{w},\hat{S}] \;\;=\;\; i\hbar,\;\;\;\;
  [\hat{\mathcal{E}},t]\;\; =\;\; i\hbar \\ \label{E1.3}
  [\bm{\hat{x}},w]&=& [S,\bm{\hat{x}}]\;\;
  =\;\;[w,\bm{\hat{\pi}}]\;\;=\;\;
  [S,\bm{\hat{\pi}}]\;\;=\;\;0.
\end{eqnarray}
while the corresponding operators are
\begin{eqnarray}\label{E1.5}
  \bm{\hat{\pi}} &=&\frac{\hbar}{i}\,\frac{\partial}{\partial\bm{x}}\,;\;\;\;\;
  \hat{S}\;\;=\;\;\frac{\hbar}{i}\,
  \frac{\partial}{\partial w}\,;\;\;\;\;
  \hat{\mathcal{E}}\;\;=\;\; i\hbar\,\frac{\partial}{\partial t}\,;
  \\ \label{E1.6}
  \bm{\hat{x}}&=&\bm{x}\,;\;\;\;\;\hat{w}\;\;=\;\;w\,.
\end{eqnarray}
where $\mathcal{E}$ is a "super-energy", which related with a common energy $E$ as $\mathcal{E}=wE $ \footnote{The same results can be obtained also in Dirac-Feynman path integral approach with replacement the common action by the super-action.}. Uncertainty relation, which correspond to these commutative relations are
\begin{eqnarray} \label{E1.8}
  \overline{\Delta \bm{x}^2}\;\overline{\Delta \bm{\pi}^2}
   &\geq& \frac{\hbar^2}{4},\;\;\; \\ \label{E1.9}
  \overline{\Delta \bm{x}^2}\;\overline{\Delta \bm{v}^2}
   &\geq& \frac{\hbar^2}{4m^2} \left\langle\frac{1}{w^2}\right\rangle.
\end{eqnarray}
where $\langle\rangle$ designates quantum mechanics averaging, while $\bm{v}$ and $m$ are velocity and mass of an object. Since $\langle\hat{w}\rangle(t)$ can be small or large in different time (see Appendix \ref{ApQM}), the extended QM describes the objects which can demonstrate both classical and quantum behaviour in different time intervals. Moreover, in according with (\ref{E1.9}), if $m\langle\hat{w}\rangle$ becomes smaller then contemporary mass of a subatomic particle, even massive body may lost well defined trajectory and becomes a quantum object. Until now such objects were not observed, but it is unclear \textit{a priori} if it means that Hamiltonians of the real physical systems are independent on the action, or this dependence is so weak that observation of such behaviour requires cosmological time intervals.

\section{Evolution of Universe \label{UnEv}}

We could try to estimate possible dependence of Hamiltonian on the action by applying S-Hamiltonian QM to evolution of Universe. Consider a non-relativistic massive body, which is moving in the homogeneous and isotropic universe with weak gravitational potential. In the first approximation with respect to the action, corresponding S-Hamiltonian can be written in as (SI)
\begin{equation}\label{E2.1}
  \mathcal{\hat{H}}=\frac{\bm{\hat{\pi}}^2}{2m\hat{w}}-
  \frac{4\pi Gm\hat{w}}{3}\hat{\rho}\bm{\hat{x}}^2
  +\frac{\gamma}{2}\left(\hat{w}\hat{S}+\hat{S}\hat{w}\right)
\end{equation}
where $\bm{\hat{x}}$ is coordinates of a body, $\hat{\rho}$ is density of matter, $G$ is the gravitational constant, $\gamma$ is a constant and we have taken into account that $\hat{S}$ and $\hat{w}$ do not commute. In the Heisenberg representation equations of motion are
\begin{eqnarray}\label{E2.2}
  \frac{d\bm{\hat{x}}}{dt} &=&
  \frac{i}{\hbar}\,[\mathcal{\hat{H}},\bm{\hat{x}}]\;\;=
   \;\;\frac{\bm{\hat{\pi}}}{m\hat{w}}\,;\;\;\;\;\;\;\;\;
   \frac{d\bm{\hat{\pi}}}{dt}\;\;=\;\;
   \frac{i}{\hbar}\,[\mathcal{\hat{H}},\bm{\hat{\pi}}]\;\;=
   -\frac{4\pi Gm\hat{w}}{3}\hat{\rho}\bm{\hat{x}}\,; \\ \label{E2.4}
  \frac{d\hat{w}}{dt} &=& \frac{i}{\hbar}\,[\mathcal{\hat{H}},\hat{w}]\;\;
  =\;\;-\gamma \hat{w}.
\end{eqnarray}
The last equation can be solved immediately:
\begin{equation}\label{E2.5}
  \langle\hat{w}\rangle = \langle\hat{w}_0\rangle
  \exp(-\gamma t).
\end{equation}
so average of the AC-variable is decreasing with time.

Numerical solution of Eqs.(\ref{E2.2})-(\ref{E2.4}) in mean-field approximation: $\hat{\rho}\to\langle\rho\rangle=\rho(\langle R\rangle)$ (where $\hat{R}$ is distance from origin to border of matter), is shown in Fig.~1 and Fig.~2 (see Appendix \ref{ApUnEvl} for details). For density of matter less that the critical density  \footnote{$\rho_c\sim3.6\cdot10^{-27}$, which is about half of the critical density that is predicted by General Relativity. Density of visible matter in the Universe is estimated at $\rho\sim3\cdot10^{-28}\ll\rho_c$} $\langle\hat{\rho}\rangle<\rho_c$, where
\begin{equation}\label{E2.6}
  \rho_c\simeq \frac{3\gamma^2}{16\pi G}
\end{equation}
the theory predicts accelerated expansion of the matter (without assumption about existing of a dark matter or dark enetgy), with deceleration parameter is between to $-1.2<q<0$ that is in a qualitative agreement with observable value $q\sim -1.08\pm0.29$. Velocity of the expansion demonstrates good agreement with the Hubble law (Fig.~1) with Hubble parameter $0.7\gamma\leq H\leq \gamma$. Therefor we can estimate $\gamma$ as $\gamma\sim H_0\simeq 2\cdot10^{-18}s^{-1}$. This means that dependence of S-Hamiltonian on the action is very weak and should be taken into account only for "cosmological" time intervals that are in order of billions years.

It can be seen in the Fig.~1 and Fig.~2
\begin{figure}
  \centering
  \includegraphics[width=15cm]{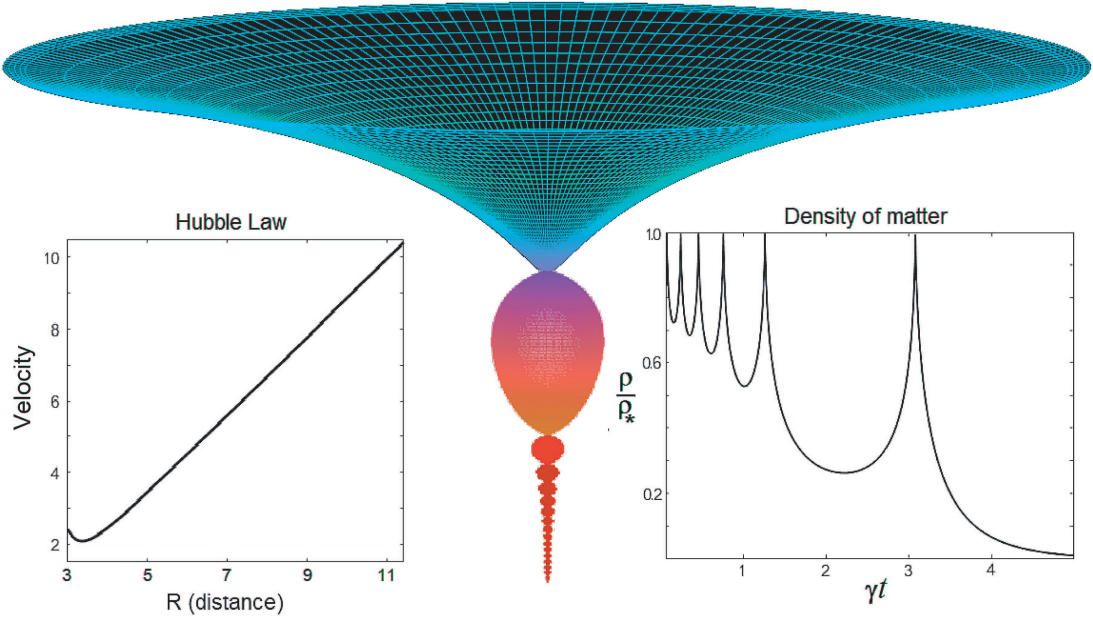}
  \caption{Evolution of Universe, Hubble Law and dynamics of matter density ($\rho_*$ is initial density). For sake of visibility the gray picture shows numerical solution for 2D Universe, while evolution of radius of 3D Universe is shown in Fig.~2. All solutions are found by using "MatLab 2020b" software.}
\end{figure}
\begin{figure}
\centering
  \includegraphics[width=14cm]{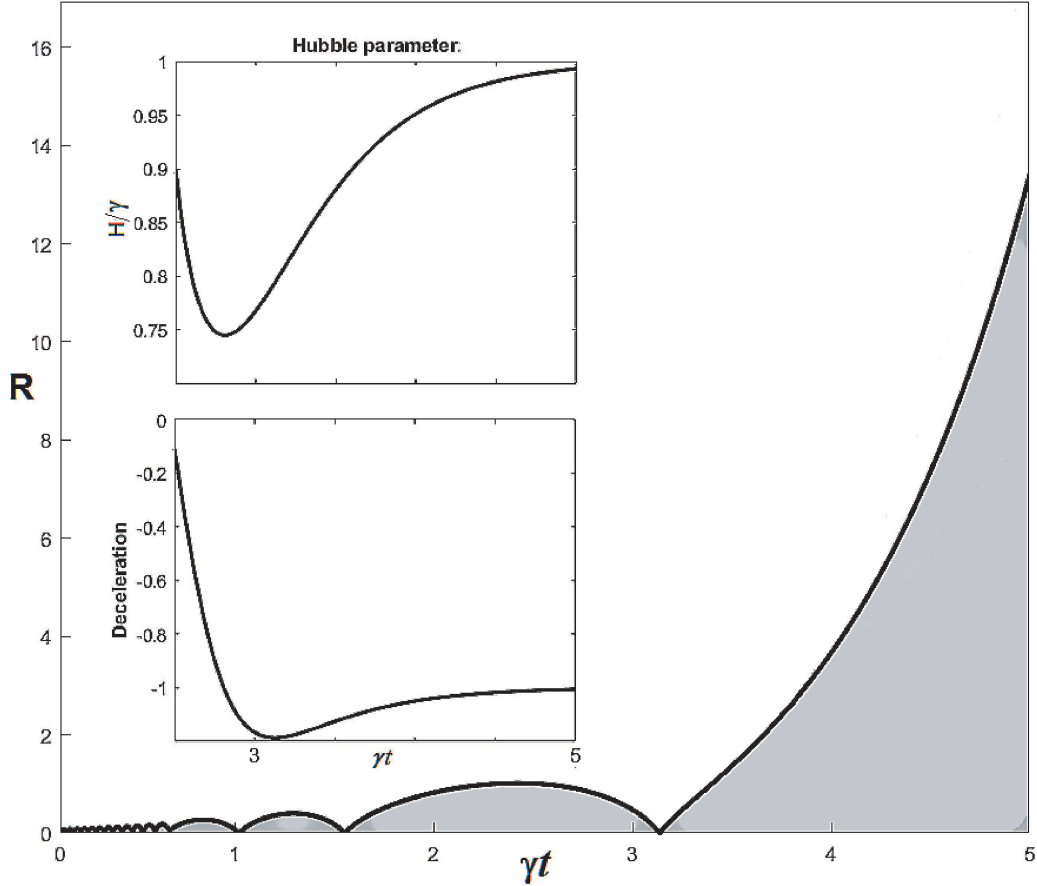}
  \caption{Dynamics of matter, Hubble parameter $H(t)$ and deceleration parameter $q=-1-\dot{H}/H^2$.}
\end{figure}
that for initial density of matter that is more than the critical density (\ref{E2.6}), the last "Big Bang", where our open accelerated universe was created, should be preceded by set of the closed universes. Duration of this process can be roughly estimated as
\begin{equation}\label{E2.9}
  t_*\sim \frac{1}{6\gamma}\ln\left(\frac{\rho_{int}}{\rho_c}\right)
\end{equation}
where $\rho_{int}$ is initial density of the universe, and could be much lager then commonly accepted age of Universe, while number of these universes can be large
\begin{equation}\label{E2.9a}
  N \sim \sqrt{\frac{\rho_{int}}{\rho_c}},
\end{equation}

In spite of such behaviour resemble the Inflation theory, there is important difference. In the Inflation theory the precursor universes are result of the thermodynamics fluctuations and could exist only a short time. On the other hand, in S-Hamiltonian dynamics existence the precursor universes is entirely dynamical effect and could require a long time.

In according with considered extension of quantum mechanics, quantization of  matter can be evolutionary process. Therefor, it is possible that in the first universes even elementary particles were almost classic objects, while in a far future $t\gg t_*$ even  stars and planets, would demonstrate quantum behaviour.

Solution of the Kepler problem, where stars are rotating surround very massive nucleus, predicts formation of the spiral galaxies, where the rotation velocity of the stars almost independent on distance for the large distances from nucleus (see Fig.~3 and Appendix \ref{ApKepler}).

\begin{figure}
\centering
  \includegraphics[width=14cm]{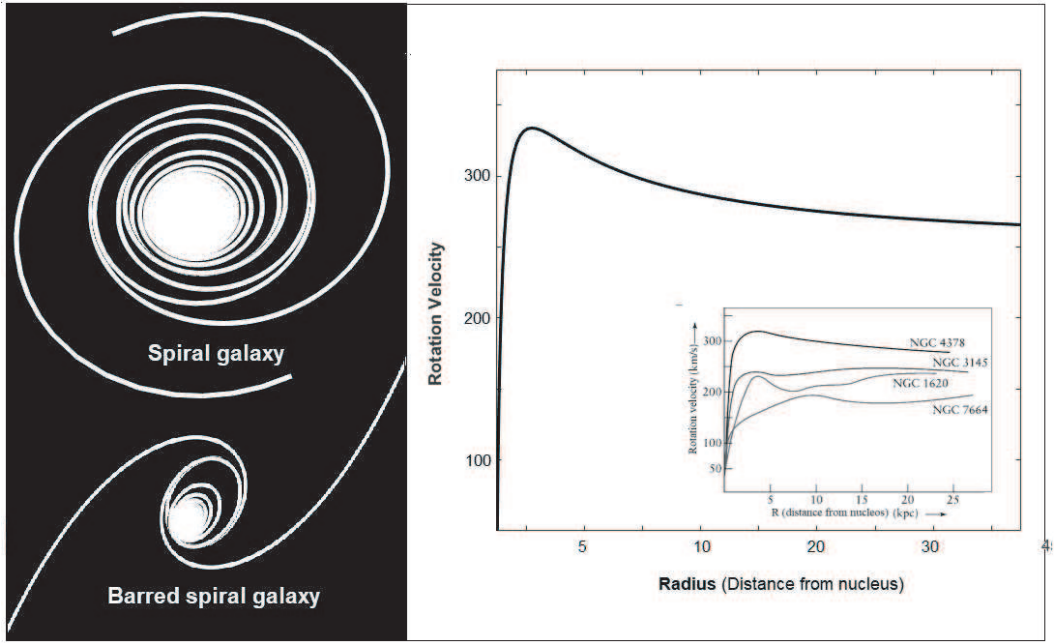}
  \caption{Formation of spiral galaxies (left) and their rotation velocity. Barred spiral galaxies correspond to the older galaxies or to galaxies with less massive nucleolus.  Right-down corner - experimental data (adopted from \cite{RotWeb}).}
\end{figure}

On the other hand, classic Kepler problem leads to dependence $V_{rot}\sim R^{-1/2}$. Independence of the rotation velocity on the large distances from a nucleolus  was observed for the most spiral galaxies \cite{Roten}. Commonly, difference between classical Kepler theory and observation is explained by the dark matter-halo effect, while the S-Hamiltonian dynamics does not need this hypothesis.

\section{Discussion}\label{Diss}

It follows from Eqs.(\ref{E1.9}) and (\ref{E2.5}) that the "position-velocity" uncertainty
\begin{equation} \label{ED.0}
  \overline{\Delta \bm{x}^2}\;\overline{\Delta \bm{v}^2}
   \geq\frac{\hbar^2}{4m^2} \left\langle\frac{1}{w^2}\right\rangle=
   \frac{\hbar^2}{4m^2}\left\langle
   \frac{1}{w^2_0}\right\rangle\,e^{2\gamma t}
\end{equation}
increases with time.  Therefore, even massive bodies will lose well-defined trajectories in the distant future and become quantum-like objects (evolutionary quantization). However, because of the small value of $\gamma$, this process would require cosmological time intervals.

As we see in Section \ref{UnEv}, the first approximation of the S-Hamiltonian with respect to action explains the accelerated expansion of the Universe, Hubble’s law, and formation of spiral galaxies with a non-Kepler curve of rotation velocity, without considering the existence of dark energy. Alternatively, the proposed theory can be interpreted as a phenomenological description of the influence of dark energy on universe dynamics.

It should be emphasized, that Eqs. (\ref{E2.2})-(\ref{E2.4}) is not adequate for high density of matter, where both mean-field approximation and weakness of the gravitational field are incorrect. Therefore, in vicinity of the collapse points our results could be only qualitative. On the other hand, since a system spends short time near these points, the AC-variable can be well approximated as a constant. So, the ordinary quantum mechanics and the General Relativity should be adequate there, with accuracy $o(\gamma \Delta t)$ \footnote{While we have considered only non-relativistic systems in this paper,  the proposed theory is relativistic-covariant, which can be seen by rewriting the basic equation (\ref{EQ2.1}) as
$$
  cu^k_{(m4)}\nabla_{(4)k}\,\mu + \frac{L}{\bar{\gamma}}\frac{\partial\mu}{\partial S}=0,
$$
where $c$ is velocity of light and
$$
   \bm{u}_{(m4)}=\frac{1}{\bar{\gamma}}\left(1,\frac{\bm{v}_m}{c}
   \right),\;\;\;\;
   \nabla_{(4)}=\left(\frac{1}{c}
   \frac{\partial}{\partial t},\nabla\right),\;\;\;\;
   \bar{\gamma}=\sqrt{1-v^2_m/c^2},
$$
are the 4-velocity and 4-gradient.}.

The proposed theory provides a simple explanation for the observable asymmetry between the matter and antimatter distributions. In accordance with CPT invariance, we should assume that for antimatter, similar to charge, mass and energy, the sign of $\gamma$ must be opposite:
\begin{equation}\label{EQD.1}
  \gamma_{antimat}\rightarrow -\gamma_{mat}.
\end{equation}
However, a negative $\gamma$ leads to a decelerated contraction instead of an accelerated expansion. Therefore, even if the matter-antimatter distributions are fully symmetrical at the beginning of the universe, they have different evolution. When the matter distribution becomes spread out due to expansion, the antimatter distribution becomes compact due to contraction This would indicate that the small quantities of antimatter that we now observe mainly consist of newborn antiparticles (and, perhaps, a small amount of the antiparticles with high initial speed), while the main part of antimatter is concentrated near the origin of the universe.

It should be noted that although the dependence of S-Hamiltonians on the action was obtained from general arguments, universality of $\gamma$ is only a simple assumption . Generally, in multi-systems, each subsystem has its own Hamiltonian and action, and thus, could have its own $\gamma_i$  \footnote{Apparently, such situations is common for living systems \cite{s-ManyBody}. Moreover, in general case, $\gamma(\bm{x},\dot{\bm{x}},t)$ could be functions of space, time and velocity.}. However, it was shown in \cite{s-ManyBody} that such a situation leads to the inevitable formation of dominance hierarchies, where the subsystem with the largest $\gamma_D=max(\gamma_i)$ becomes dominant, while all other subsystems lose individuality and follow the dominant system. Because this situation  is not common for physical system behavior, the assumption about the universality of $\gamma$ is, apparently, a reasonable hypothesis.

It is important to experimentally verify that the action dependence of the Hamiltonian is actually realized in real systems. At first glance, because of evolutionary quantization, we could observe differences in the light emission of old and young stars. However, because the difference in the energy spectra of the atoms appears only in the second approximation with respect to $\gamma$ (see Appendix \ref{ApQM}), it is unobservable because of the small value of $\gamma$.

Unlike the energy spectrum, the spatial characteristics of the atoms acquire a time-dependent scale factor in the first approximation with respect to $\gamma$  (see Appendix \ref{ApQM}). For example, in extended quantum mechanics, the size of an atom can be estimated as
\begin{equation}\label{EQD.2}
  \langle r\rangle=e^{2\gamma t}\langle r\rangle_0 +o(\gamma),
\end{equation}
where $\langle r\rangle_0$ is the atom size in ordinary quantum mechanics. Unfortunately, even for observations over a year, the corresponding contribution will be $10^{-11}-10^{-10}$, which is much smaller than the available accuracy of $10^{-3}$.

There is, however, a field, where we  is not restricted by the small size of the aforementioned effects.  Given that in \cite{ s-langr}, \cite{ sndlr-book} derivation of the S-Hamiltonian equations (see Sec. \ref{ApSetUp} ) did not contain specific properties of the describing systems and could describe both reversible and non-reversible dynamics. S-dynamics can be applied to the description of non-physical systems as well.

About 100 years ago, Hans Selye \cite{Selye}, who discovered stress, and  Walter Cannon \cite{Canon}, who discovered homeostasis, proposed that the primary motivation for living species’ activity is an attempt to cope with stress (see \cite{sndlr-tsitol} for a comprehensive discussion). Using stress as an analog of  "action" in physical systems, we can apply S-Hamiltonian dynamics to describe the behavior of living systems. This approach leads to a reasonable description of a wide class of phenomena, from bacterial chemotaxis to dynamics of homeostasis, the appearance of hierarchy in social  groups, dynamics of drug addiction, and behavior of living species in the environment \cite{s-ManyBody} - \cite{sndlr-book}. Moreover, in generalized quantum mechanics, objects that can demonstrate both quantum and classical behavior that reminds the well-known property of consciousness  \cite{LotkaQB}, \cite{Vannini2008}. Therefore, this approach could be a useful theoretical tool for the rational description of brain function (see \cite{sndlr-book} for details).

\appendix

\section{Set up of the approach} \label{ApSetUp}

Consider a system moving in a space with coordinates $\bm{x}=\{x_1,...,x_n\}$. We assume that it is not possible to obtain the exact value of the coordinates $x_i$ or the velocities $v_i$. It can only be stated that there is some \textit{possibility} that at time $t$, the system is close to the point $\bm{x}$, and its velocity is close to $\bm{v}$. In such a case, the movement of the system can be described as follows: if we denote the possible values of the velocity $\bm{v}$ as $\bm{v}',\bm{v}'',\bm{v}''',...$ we can state that
\begin{itemize}\label{D1}
    \item If the system is in the vicinity of the
    point $\bm{x}$ at the time $t+dt$, then at the previous time $t$, the system could have been near the point
    $\bm{x}'\approx \bm{x}-\bm{v}'dt$, or
    $\bm{x}''\approx \bm{x}-\bm{v}''dt$, or      $\bm{x}'''\approx \bm{x}-\bm{v}'''dt$, or ..., and so on, for all possible
    values of the velocity $\bm{v}$,
\end{itemize}
that, actually, reflects \textit{causality principle} for dynamics of the systems. Let us denote the \textit{possibility} that the system is in a small domain $\Delta_{\bm{x}}$ around the point $\bm{x}$ at the time $t$ as $m(\Delta_{\bm{x}},t)$. We denote the \textit{possibility} that in a domain $\Delta_{\bm{x}}$ and at the time $t$, the system velocity is in a small domain $\Delta\bm{v}$ by $P(\Delta\bm{v}|\Delta_{\bm{x}},t)$. Then, the preceding expression can be symbolically written as follows:
\begin{eqnarray} \nonumber
    m(\Delta_{\bm{x}},t+dt) &=& \mathcal{C}\left\{ T\left[P(\Delta\bm{v}'|\Delta_{\bm{x}'},t);
    m(\Delta_{\bm{x}'},t)\right]; ... \right. \\ \nonumber
    & & \left. ...;T\left[P(\Delta\bm{v}''|\Delta_{\bm{x}''},t);
    m(\Delta_{\bm{x}''},t)\right]; ... \right. \\ \nonumber
    & & \left. ...T\left[P(\Delta\bm{v}'''|\Delta_{\bm{x}'''},t);
    m(\Delta_{\bm{x}'''},t)\right];...\right. \\ \label{EQ1.13w}
    & & ... \left. \mbox{ and so on}\right\},
\end{eqnarray}
where $\mathcal{C}\left\{...;...\right\}$ and $T[...;...]$ are symbolic expressions for the logical connectives ``\textbf{OR}" and ``\textbf{AND}". In fact, (\ref{EQ1.13w}) is simply the previous natural language expression, written in symbolic form. To translate (\ref{EQ1.13w}) into an equation, we should define the \textit{mathematical representations} of the symbols $\mathcal{C}\left\{...;...\right\}$ and $T[...;...]$ and $m(\Delta_{\bm{x}},t)$ and $P(\Delta\bm{v}|\Delta_{\bm{x}},t)$.

The common candidates for mathematical representations of the logical connectives are the various triangular norms and conorms \cite{TriangNorm}. A triangular conorm is a binary operation that satisfies of the properties of common logic. Namely, the connective $OR$ is represented by a binary operation $\mathcal{C}\{\mu_1,\mu_2\}$ that satisfy the following:
\begin{eqnarray}\label{EQ1.2}
    & &  \mathcal{C}\{\mu_1,\mu_2\} = \mathcal{C}\{\mu_2,\mu_1\}, \\ \label{EQ1.3}
    & &  \mathcal{C}\{\mu, 0\} = \mu,;\;\ \mathcal{C}\{\mu, 1\} = 1, \\ \label{EQ1.4}
    & &  \mbox{If } \mu_2\leq\mu_3, \mbox{ then }
    \mathcal{C}\{\mu_1,\mu_2\}\leq \mathcal{C}\{\mu_1,\mu_3\},
\end{eqnarray}
where condition (\ref{EQ1.2}) reflects the symmetry of the ``\textit{OR}" connective, while condition (\ref{EQ1.4}) reflects its  monotonicity.

Unfortunately, there is an infinite number of triangular conorms $\mathcal{C}\left\{\mu_1;\mu_2\right\}$, which is not ideal for a robust theory. It is remarkable, however, that the natural properties of the local topology of the state space drastically restricts the available choices for the representation of this connective. To demonstrate this characteristic, let us consider two nearest-neighbor domains $\Delta_1$ and $\Delta_2$ of the system state space. It is evident that the possibility that the system is in the joint domain $\Delta_1\bigcup\Delta_2$ is equal to the possibility that it is in the domain $\Delta_1$ or it is in the domain $\Delta_2$. Hence, we can state the following:
\begin{equation*}
    m\left(\Delta_1\bigcup\Delta_2\right) =
    \mathcal{C}\left\{m(\Delta_1); m(\Delta_2)\right\}.
\end{equation*}
If both domains are collapsed to the same point $\Delta_{1},\Delta_{2} \to \bm{x}$, we have
\begin{equation*}
  m(\Delta_1)\to m(\Delta_2)\to m\left(\Delta_1\bigcup\Delta_2\right) \to \mu(\bm{x}),
\end{equation*}
which implies that
\begin{equation}\label{EQ1.15y}
    \mathcal{C}\left\{\mu(\bm{x});\mu(\bm{x})\right\}=\mu(\bm{x}).
\end{equation}
It is shown in \cite{s-langr} that this condition leads to a unique representation of the connective $\mathcal{C}\left\{\mu_1;\mu_2\right\}$:
\begin{equation}\label{EQ1.11}
    \mathcal{C}\{\mu_1,\mu_2\} = \max\{\mu_1\;\mu_2\}.
\end{equation}

It is understood that $m(\Delta_{\bm{x}})$ should correspond to a particular measure of the domain $\Delta_{\bm{x}}$. If we are interested in distances much greater than the typical size of the domains $\Delta_{\bm{x}}$, it is reasonable to consider a limit where the domain collapses to a point:
\begin{equation}\label{EQ1.14x}
    \lim_{\Delta_{\bm{x}}\to\bm{x}}m(\Delta_{\bm{x}},t) = \mu(\bm{x},t).
\end{equation}
Theoretically, there are two cases
\begin{eqnarray} \label{EQ1.14a}
    \mu(\bm{x},t) &\equiv& 0, \\ \label{EQ1.14b}
    \mu(\bm{x},t) &\neq& 0
\end{eqnarray}
where $\mu(\bm{x},t)$ is the possibility that at the time $t$ the system is at the point $\bm{x}$. The first case is equivalent to the probabilistic approach to dynamical problems, where $\mu(\bm{x},t)\Rightarrow\varrho(\bm{x},t)dV$ and $\varrho(\bm{x},t)$ can be identified with probability density. The second case, which  corresponds to the so-called \textit{atomic measure,} was not applied to  dynamical problems previously \footnote{It should be emphasized that the possibility function $\mu(\bm{x},t)$ cannot be identified with any probability density, because they have different mathematical features.  Possibility function is pointwise limited  $0\leq\mu(\bm{x},t)\leq 1$, while the integral of it over all space could be infinite. On the other hand, integral of probability density over all space is limited, while probability density itself can be infinite at some points. Actually, possibility is described by a \textit{function}, while probability is described by a \textit{functional}}.

We assume that $\mu(\bm{x},t)$ and $P(\bm{v};\bm{x},t)$, where $P(\bm{v};\bm{x},t)$ is the possibility that for the point $\bm{x}$ and time $t$, the value of the velocity is $\bm{v}$,  are continuous, bounded functions $0\leq \mu,P\leq 1$, where the values $0$ and $1$ correspond to the minimal and maximal possibilities, respectively. It is assumed also that an infinite velocity is impossible, therefore, $P(\pm\infty;\bm{x},t)=0$.

Using (\ref{EQ1.14x}), we can rewrite (\ref{EQ1.13w}) as
\begin{equation}\label{EQ1.17}
    \mu(\bm{x},t+dt) = \sup_{\bm{v}}{T[P(\bm{v};\bm{x},t);\mu(\bm{x}-\bm{v}dt,t)]},
\end{equation}
If the primary concern is the time intervals, which are much more than $dt$, it is reasonable to take the limit $dt\to 0$. Let us designate the velocity corresponding to the maximal value of the right side of (\ref{EQ1.17}) as $\bm{v}_m$. Thus, we can state that
\begin{equation}\label{EQ1.19}
    \sup_{\bm{v}}T[P(\bm{v};\bm{x},t);\mu(\bm{x}-\bm{v}dt,t)] = T[P(\bm{v}_m;\bm{x},t);\mu(\bm{x}-\bm{v}_m dt,t)]
\end{equation}
Given that the function $T[P,\mu]$ is monotonically increasing, $T[P,\mu]\leq\min(P,\mu)\leq\mu$ and $T[1,\mu]=\mu$, it can be shown \cite{s-langr},\cite{sndlr-book} that for a given $\mu$, the maximum of $T[P,\mu]$ should be equal to $\mu$. Therefore
\begin{equation}\label{EQ1.19a}
    T[P(\bm{v}_m;\bm{x},t);\mu(\bm{x}-\bm{v}_m dt,t)] = \mu(\bm{x}-\bm{v}_m dt,t)
\end{equation}
Thus, it follows from (\ref{EQ1.17}) that
\begin{equation}\label{EQ1.20}
 \mu(\bm{x},t+dt)= \mu(\bm{x}-\bm{v}_m dt,t).
\end{equation}

For a small $dt$, we can expand $\mu(\bm{x}-\bm{v}_mdt,t)$ with respect to $dt$ and in the limit $dt\to 0$, we obtain
\begin{equation}\label{EQ1.21}
    \frac{\partial\mu}{\partial t} + (\bm{v}_m\cdot\nabla\mu) = 0.
\end{equation}
It should be noted that for $\mu(\bm{x},t)-(\bm{v}_m\cdot\nabla\mu)dt$ to be maximal, $(\bm{v}_m\cdot\nabla\mu)$ should be minimal. On the other hand, it follows from (\ref{EQ1.19a}) that for $dt\equiv0$
\begin{equation}\label{EQ1.22}
    T[P(\bm{v}_m;\bm{x},t);\mu(\bm{x},t)]=\mu(\bm{x},t).
\end{equation}
Therefore, $\bm{v}_m(\mu,\nabla\mu;\bm{x},t)$ can be found by minimization of
\begin{equation}\label{EQ1.23x}
    (\bm{v}_m\cdot\nabla\mu)\to \min,
\end{equation}
under the restriction
\begin{equation}\label{EQ1.24x}
    P(\bm{v}_m;\bm{x},t)= \zeta_T(\mu),
\end{equation}
where $\zeta_T$ is a solution of the equation $T[\zeta;\mu]=\mu$.

The solution of the system (\ref{EQ1.23x}),(\ref{EQ1.24x}) is well-known and can be solved by the method of \textit{Lagrange multipliers}
\begin{eqnarray}\label{EQ1.26x}
  \lambda\frac{\partial P}{\partial \bm{v}_m} &=& \nabla\mu, \\ \label{EQ1.26a}
  P(\bm{v}_m;\bm{x},t) &=& \zeta_T(\mu),
\end{eqnarray}
where $\lambda>0$, such that $\bm{v}_m$ corresponds to the minimum of (\ref{EQ1.23x}).

The system of equations (\ref{EQ1.21}),(\ref{EQ1.26x}), and (\ref{EQ1.26a}) possess several important features. First, the region of the most possible ($\mu(\bm{x}(t),t)=1$) and impossible ($\mu(\bm{x}(t),t)=0$) trajectories do not depend on concrete representations of the connective $T[P,\mu]$ because $T[P,1]=P$ and $T[P,0]=0$ for any $T[P;\mu]$. In addition, the most possible trajectories do not depend on subjectivity in the assignment of intermediate values of possibility for the states of the initial system. Therefore, these trajectories contain the most reliable information about the system's behavior, and only this case should be considered to the description of real-world problems.

Consider a dynamical system, whose behavior is described by $D+1$ variables: $D$ features that we will consider as ``coordinates" $\bm{x}$ in the system's state space and an additional scalar variable $S$ (\textit{S-variable}). Let us designate $\bm{v}$ as the velocity of the system "movement" in D-dimension space and $L$ as the rate of change of the $S$-variable. Therefore, $\bm{v}_m$ in (\ref{EQ1.21}) takes the form $\bm{v}_m=\{\bm{v}, L\}$, where
\begin{eqnarray*}
  \bm{v} &=& \frac{d\bm{x}}{dt}, \\
  L &=& \frac{dS}{dt}.
\end{eqnarray*}

If our knowledge of the location and velocity of the system is imprecise. The system dynamics should be described by its \textit{possibility function}
\begin{equation}\label{EQ2.0}
    \mu=\mu(S,\bm{x},t)
\end{equation}
and by the possibility function for the dynamics laws
\begin{equation}\label{EQ2.0a}
    P=P(L,\bm{v};S,\bm{x},t)
\end{equation}
In this case, equations (\ref{EQ1.21}) and (\ref{EQ1.26x})-(\ref{EQ1.26a}) take the form
\begin{equation}\label{EQ2.1}
    \frac{\partial\mu}{\partial t} + (\bm{v}\cdot\nabla\mu) + L\frac{\partial\mu}{\partial S}= 0
\end{equation}
and
\begin{eqnarray}\label{EQ2.2a}
  \lambda\frac{\partial P}{\partial \bm{v}} &=& \nabla\mu, \\ \label{EQ2.2b}
  \lambda\frac{\partial P}{\partial L} &=& \frac{\partial \mu}{\partial S}, \\ \label{EQ2.2cx}
  P(\bm{v},L;\bm{x},S,t) &=& \zeta_T(\mu).
\end{eqnarray}
We can solve Eq.(\ref{EQ2.2cx}) with respect to $L$ to obtain the following:
\begin{equation}\label{EQ2.3}
    L=L(\bm{v},\bm{x},S,\mu,t).
\end{equation}
Now, substituting (\ref{EQ2.3}) in (\ref{EQ2.0a}) and differentiating with respect to $\bm{v}$, we obtain the following:
\begin{equation*}\nonumber
   \frac{\partial P}{\partial L}\frac{\partial L}{\partial \bm{v}} +  \frac{\partial P}{\partial\bm{v}}  = 0.
\end{equation*}
Using (\ref{EQ2.2a}),(\ref{EQ2.2b}), we obtain
\begin{equation}\label{EQ2.5}
    \frac{\partial L}{\partial \bm{v}}=-\frac{\nabla\mu}{\partial_s\mu}.
\end{equation}
The solution of (\ref{EQ2.5}) with respect to $\bm{v}$ gives
\begin{equation*}
  \bm{v}=\bm{v}\left(-\frac{\nabla\mu}{\partial_s\mu},\bm{x},S,\mu,t \right).
\end{equation*}
Finally, substituting (\ref{EQ2.5}) in (\ref{EQ2.1}), we obtain
\begin{equation}\label{EQ2.6}
    \frac{\partial\mu}{\partial t} - H\left(-\frac{\nabla\mu}{\partial_s\mu},\bm{x},S,\mu,t\right) \frac{\partial\mu}{\partial S}= 0,
\end{equation}
where
\begin{equation}\label{EQ2.7}
    H=\left(\bm{v}\cdot\frac{\partial L}{\partial \bm{v}} \right) - L.
\end{equation}
Equation (\ref{EQ2.6}) is a first-order partial differential equation that can be solved using the method of characteristics. The characteristics of equation (\ref{EQ2.6}) are determined as follows:
\begin{equation}\label{EQ2.8}
  dt = \frac{d\bm{x}}{\frac{\partial\mathcal{H}}{\partial \bm{\pi}}}=-\frac{dS}{\frac{\partial\mathcal{H}}{\partial w}}=
  -\frac{d\bm{\pi}}{\frac{\partial\mathcal{H}}{\partial \bm{x}}
  -\bm{\pi}\frac{\partial\mathcal{H}}{\partial \mu}} =
  \frac{dw}{\frac{\partial\mathcal{H}}{\partial S}
  +w\frac{\partial\mathcal{H}}{\partial \mu}}=\frac{d\mu}{\frac{\partial\mu}{\partial t}-\mathcal{H}},
\end{equation}
where $\mathcal{H}=wH$ and
\begin{eqnarray}\label{EQ2.8a}
 \bm{\pi} &=& -\nabla\mu \\ \label{EQ2.8b}
        w &=& \partial_s\mu.
\end{eqnarray}
In this appear, we refer to $\bm{\pi}$ and $w$ as an \textit{S-momentum} and \textit{AC-variable}, respectively.
Transforming $\bm{\pi}$ and $w$ as
\begin{eqnarray} \label{FN1.1a}
   w &\rightarrow& \phi(t)w, \\  \label{FN1.1b}
   \bm{\pi} &\rightarrow& \phi(t)\bm{\pi}
\end{eqnarray}
where $\phi(t)$ satisfies
\begin{equation}\label{FN1.3}
  \frac{d\phi}{dt}=\phi\frac{\partial \mathcal{H}}{\partial \mu}.
\end{equation}
we can write a system of ordinary differential equations, which corresponds to the characteristics (\ref{EQ2.8}) as:
\begin{eqnarray} \label{EQ2.9r}
  \frac{d\bm{x}}{dt} &=& \frac{\partial\mathcal{H}}
  {\partial \bm{\pi}}, \\  \label{EQ2.9xr}
  \frac{d\bm{\pi}}{dt} &=& -\frac{\partial\mathcal{H}}{\partial \bm{x}},
  \\  \label{EQ2.9xxxr}
  \frac{dw}{dt} &=& \frac{\partial\mathcal{H}}{\partial S},
  \\ \label{EQ2.9xxr}
  \frac{dS}{dt} &=& -\frac{\partial\mathcal{H}}{\partial w}\;\;=\;\;L.
\end{eqnarray}
If $P(L,\bm{v};\bm{x},t)$ does not explicitly depend on $S$, then $L(\bm{v},\bm{x},\mu_0,t)$ and $H(\bm{p},\bm{x},\mu_0,t)$ do not depend on $S$, either. In this case $w\equiv1$, so
$\bm{\pi}=\bm{p},\;\mathcal{H}=H$ become an ordinary momentum and an ordinary Hamiltonian, while equations (\ref{EQ2.9r})-(\ref{EQ2.9xxr})  become the well-known Hamiltonian equations of the classical mechanics.

The S-Hamiltonian $\mathcal{H}=\mathcal{E}$ can be considered as the  generalized energy because it is conserved along the system trajectories, while the common mechanical energy  $H=E$ is not conserved for the S-dependent Hamiltonians. It should be noted, that for action dependent S-Hamiltonians the dynamics equations can be time-irreversible even if the S-Hamiltonian is time-independent.

Alternatively, we can obtain S-Hamiltonian equations (\ref{EQ2.9r})-(\ref{EQ2.9xxr}) by variation of the S-action $\mathbb{S}$ with respect to $\bm{x}$, $\bm{\pi}$, $w$, $S$ and $t$
\begin{eqnarray} \nonumber
  \delta \mathbb{S}
  &=& (\bm{\pi}\cdot\delta\bm{x}) + S\delta w-\mathcal{H}\delta t \\ \label{E2.1wb}
  &+& \int_0^t\left\{-\left[\frac{dS}{dt}+
  \frac{\partial \mathcal{H}}{\partial w}\right]\delta w+
  \left[\frac{dw}{dt}-\frac{\partial\mathcal{H}}{\partial S}\right]\delta S+\right.\\ \nonumber
  &+&\left.\left(\left[\frac{d\bm{x}}{dt}-
  \frac{\partial\mathcal{H}}{\partial\bm{\pi}}\right]
  \cdot\delta\bm{\pi}\right)-\left(\left[\frac{d \bm{\pi}}{dt}+
  \frac{\partial \mathcal{H}}{\partial\bm{x}}\right]\cdot
  \delta\bm{x}\right)\right\}dt.
\end{eqnarray}

It is important that although our consideration has began from the definition of the \textit{possibility functions}, it does not need to know its explicit forms if we are interested only in the most possible behavior of a system. Instead, only expressions for S-Hamiltonian or S-Lagrangian are required.

It should be emphasized that equations (\ref{EQ2.9r})-(\ref{EQ2.9xxr}) were obtained by using only the master equation (\ref{EQ1.17}), which is logically followed from the causality principle and local topology of a state space. This implies that principle of least action, which is equivalent Eqs.(\ref{EQ2.9r})-(\ref{EQ2.9xxr}) is not an independent axiom, but, in fact, a logical consequence of the causality principle.

\section{Energy spectra and evolutionary scale factors in the extended quantum mechanics.} \label{ApQM}

To study the energy spectrum of a system in extended quantum mechanics, it is more convenient to use the Schrodinger representation.
\begin{equation}\label{EQ5.1}
  \imath\hbar\frac{\partial\Psi}{\partial t} =
  w\mathcal{\hat{H}}_0\left(\frac{\bm{\hat{\pi}}}{w}\,,
  \bm{x}\right)\Psi +\frac{\gamma}{2}(w\hat{S} + \hat{S}w)\Psi,
\end{equation}
where $\hat{H}_0$ is an ordinary (Hermitian) Hamiltonian, and the usual operator for momentum $\bm{\hat{p}}$ is replaced by $\bm{\hat{\pi}}/w$
\begin{equation}\label{EQ5.2}
  \bm{\hat{p}}\rightarrow\frac{\bm{\hat{\pi}}}{w}=
  \frac{\hbar}{\imath w}\nabla,
\end{equation}
$\hat{S}$ is operator of action
\begin{equation}\label{EQ5.3}
  \hat{S}=\imath\hbar\frac{\partial}{\partial w},
\end{equation}
 we have considered that $w$ and $\hat{S}$ do not commute. It is reasonable to search for a solution in the form
\begin{equation}\label{EQ5.4}
  \Psi(\bm{x},w,t)=\left[\exp{-\frac{\imath}{\hbar}\,
  \mathcal{E}}(w,t)\right]\varphi(\bm{x},w),
\end{equation}
Substituting (\ref{EQ5.4}) into (\ref{EQ5.1}), we split this equation,
\begin{eqnarray} \label{EQ5.5}
   \frac{\partial\mathcal{E}}{\partial t}-
   \gamma w\frac{\partial\mathcal{E}}{\partial w}
   -\imath\frac{\hbar\gamma}{2}&=&w\epsilon(w),  \\ \label{EQ5.5a}
   \mathcal{\hat{H}}_0\left(\frac{\bm{\hat{\pi}}}{w}\,,
  \bm{x}\right)\varphi+\imath\hbar\gamma
  \frac{\partial\varphi }{\partial w}&=&\epsilon(w)\varphi,
\end{eqnarray}
with an unknown function $\epsilon(w)$ that should be determined from the boundary conditions.

In general, $\epsilon(w)$ is a complex-valued function, but we can eliminate its imaginary part, $\epsilon_I(w)=\mathrm{Im}\,\epsilon(w)$, by replacing 
\begin{eqnarray}\label{EQ5.6}
  \varphi &\rightarrow& \exp\left(\frac{1}{\hbar\gamma}
 \int_0^w dw'\epsilon_I(w')\right)\phi(\bm{x},w),
  \\ \label{EQ5.6a}
  \mathcal{E} &\rightarrow& \mathcal{E} -
  \frac{\imath}{\gamma}\int_0^w dw'\epsilon_I(w'),
\end{eqnarray}
which leads to
\begin{eqnarray} \label{EQ5.7}
   \frac{\partial\mathcal{E}}{\partial t}-
   \gamma w\frac{\partial\mathcal{E}}{\partial w}
   -\imath\frac{\hbar\gamma}{2}&=&w\epsilon_R(w),  \\ \label{EQ5.7a}
   \mathcal{\hat{H}}_0\left(\frac{\bm{\hat{\pi}}}{w},
  \bm{x}\right)\phi+\imath\hbar\gamma
  \frac{\partial\phi}{\partial w}&=&\epsilon_R(w)\phi,
\end{eqnarray}
Therefore, only the real part of the function $\epsilon(w)$, $\epsilon_R=\mathrm{Re}\,\epsilon(w)$, should be determined. Combining (\ref{EQ5.7a}) with its complex-conjugate equation, we obtain,
\begin{equation}\label{EQ5.8}
  \frac{\partial|\phi|^2}{\partial w} +
  \frac{\imath}{\hbar\gamma}
  \left(\phi \mathcal{\hat{H}}_0^*\phi^* -
   \phi^* \mathcal{\hat{H}}_0\phi \right)=0,
\end{equation}
Therefor, for Hermitian $\mathcal{\hat{H}}_0$
\begin{equation}\label{EQ5.9}
  \int_V d^3\bm{x}|\phi(\bm{x},w)|^2=const.=1,
\end{equation}
is independent on $w$ and can be normalized to one.

The solution of Eq.(\ref{EQ5.7}) is
\begin{equation}\label{EQ5.10}
  \mathcal{E}=-\frac{1}{\gamma}\int_0^w dw'\epsilon_R(w')+
  \imath\frac{\hbar\gamma}{2}\,t +
  F\left(we^{\gamma t}\right),
\end{equation}
where $F\left(we^{\gamma t}\right)$ is an arbitrary function
that can be determined from the condition that for $\gamma\to 0$, extended quantum mechanics must be consistent with ordinary quantum mechanics.  This implies that $\lim_{\gamma\to0}\mathcal{E}=E^{(0)}t$, where $E^{(0)}$ is the energy spectrum in ordinary quantum mechanics. Therefore, $F\left(we^{\gamma t}\right)$ and $\mathcal{E}(w,t)$ can be written as
\begin{eqnarray}\label{EQ5.11}
  F\left(we^{\gamma t}\right)&=&\frac{1}{\gamma}
  \int_0^{we^{\gamma t}} dw'\epsilon_R(w') +
  \imath\frac{\hbar}{2}\ln\Delta_\sigma
  \left(we^{\gamma t}\right)+\gamma\xi\left(we^{\gamma t}\right),  \\  \label{EQ5.11a}
  \mathcal{E}(w,t)&=&\frac{1}{\gamma}
  \int_w^{we^{\gamma t}} dw'\epsilon_R(w') +
  \imath\frac{\hbar}{2}\ln\Delta_\sigma
  \left(we^{\gamma t}\right)+\gamma\xi\left(we^{\gamma t}\right),
\end{eqnarray}
where $\xi\left(we^{\gamma t}\right)$ is an arbitrary function that provides corrections $o(\gamma^2)$ to the energy and wave functions and can be omitted in the first approximation with respect to $\gamma$.

$\Delta_\sigma\left(we^{\gamma t}\right)$ should satisfy
\begin{eqnarray}\label{EQ5.12}
  & & \lim_{\sigma\to 0}\Delta_\sigma(z)\rightarrow
  \delta(z-1), \\ \label{EQ5.12a}
  & & \int_0^\infty dz\,\Delta_\sigma(z) = 1,
\end{eqnarray}

Similar to ordinary quantum mechanics, the energy spectrum $E^{(\gamma)}$ can be obtained from
\begin{eqnarray}\nonumber
  & & E^{(\gamma)}_n = \langle\Psi_n^*|\imath\hbar\frac{\partial}
  {\partial t}\,|\Psi_n\rangle = \int_0^\infty \frac{\partial\mathcal{E}(w,t)}{\partial t}\,e^{\gamma t}dw
  \int_Vd^3\bm{x}|\phi|^2 \\ \label{EQ5.13}
  & & =\int_0^\infty dz\Delta_\sigma(z)\left[z\epsilon_R(z) + \imath\frac{\hbar\gamma}{2}+\imath\frac{\hbar\gamma}{2}z
  \frac{\Delta'_\sigma}{\Delta_\sigma}+
  o(\gamma^2)\right]\int_Vd^3\bm{x}|\phi|^2=  \\ \nonumber
  & & =\int_0^\infty dz\Delta_\sigma(z)\,\epsilon_{Rn}(z)
  =\epsilon_{Rn}(1)+o(\gamma^2),
\end{eqnarray}
where we designated $z=e^{\gamma t}w$. Since $\epsilon_{Rn}(1)=E_n^{(0)}$ we have
\begin{equation}\label{EQ5.14}
  E^{(\gamma)}_n=E_n^{(0)}+o(\gamma^2),
\end{equation}
\textit{i.e.} in extended quantum mechanics, the energy spectrum of a system matches the corresponding energy spectrum in  ordinary quantum mechanics with accuracy up to $o(\gamma^2)$.

However, in extended quantum mechanics, space characteristics can acquire a time-dependent scale factor. For example, for systems with Coulomb interactions (atoms), Eq.(\ref{EQ5.7a}) takes the following form: Eq.(\ref{EQ5.7a}) takes a form
\begin{equation}\label{EQ5.15}
  \left[-\frac{\hbar^2}{2mw^2}\nabla_i\nabla^i + \frac{1}{2}\sum_{ij} \frac{\alpha_{ij}}{|\bm{x}_i-\bm{x}_j|}  +\imath\hbar\gamma\frac{\partial }{\partial w}\right]\,\phi
  = \epsilon_R\phi,
\end{equation}
introducing a new variable $\bm{\rho}_i=w^2\bm{x}_i$ we obtain
\begin{equation}\label{EQ5.16}
  \left[-\frac{\hbar^2}{2m}\nabla_i\nabla^i + \frac{1}{2} \sum_{ij} \frac{\alpha_{ij}}{|\bm{\rho}_i-\bm{\rho}_j|} + 2\imath\frac{\hbar\gamma}{w^3}(\bm{\rho}_i\cdot\nabla^i)+ \imath\hbar\gamma\frac{\partial}{\partial w}\right]\phi
= \frac{\epsilon_R}{w^2}\,\phi(\bm{\rho},w).
\end{equation}
So, in the lowest approximation with respect to $\gamma$, the average
\begin{equation}\label{EQ5.17}
  \langle\bm{x}_i\rangle=e^{2\gamma t}\langle\bm{x}_i\rangle_0 +o(\gamma),
\end{equation}
acquires  the scale factor $e^{2\gamma t}$; however, the energy spectrum in this approximation remains the same as in the ordinary quantum mechanics.

\section{The universe expansion} \label{ApUnEvl}

We can write the Heisenberg equations of motion in the mean-field approximation (MF) as
\begin{eqnarray}\label{EQ3.1}
  \frac{d\bm{\hat{x}}}{dt} &=&
   \frac{\bm{\hat{\pi}}}{mw}, \\ \label{EQ3.2}
   \frac{d\bm{\hat{\pi}}}{dt}&=&
   -\frac{4\pi Gmw}{3}\bar{\rho}\bm{\hat{x}};
\end{eqnarray}
where $\dot{w}=-\gamma w$ and $\bar{\rho}=\langle\hat{\rho}\rangle$ is MF-average density of matter:
\begin{equation}\label{EQ3.3}
  \bar{\rho}=\frac{4\pi M}{3\bar{R}^3},
\end{equation}
where $M$ is mass of Universe and $R$ is average distance from origin to border of the matter.

Defining operator of velocity as
\begin{equation}\label{EQ3.0}
  \bm{\hat{V}}=\frac{\bm{\hat{\pi}}}{mw}=\frac{\hbar}{i}
  \frac{\nabla}{mw},
\end{equation}
we can rewrite Eqs.(\ref{EQ3.1})-(\ref{EQ3.2}) in the form
\begin{eqnarray}\label{EQ3.5}
  \frac{d\langle\bm{\hat{x}}\rangle}{dt} &=& \langle \bm{\hat{V}}\rangle, \\ \label{EQ3.6}
   \frac{d\langle \bm{\hat{V}}\rangle}{dt}&=&
   \gamma\langle \bm{\hat{V}}\rangle
   -\frac{4\pi G}{3}\bar{\rho}\langle\bm{\hat{x}}\rangle,
   \\  \label{EQ3.6f}
   \frac{d\bar{\rho}}{dt}&=&-3H\bar{\rho},
\end{eqnarray}
where we defined the Hubble parameter $H$
\begin{eqnarray}\label{EQ3.12}
  \langle \hat{V}\rangle &=& Hr,\\ \label{EQ3.0a}
  r &=&\sqrt{\langle \hat{x}_i\rangle\langle \hat{x}^i\rangle}
\end{eqnarray}
which can be found from
\begin{equation}\label{EQ3.13}
  \frac{dH}{dt}=\gamma H-H^2-\frac{\gamma^2}{4}
  \frac{\bar{\rho}}{\rho_c},
\end{equation}
This equation reminds the second Friedmann equation, where the pressure term was replaced by $-\gamma H$.

The system (\ref{EQ3.5})-(\ref{EQ3.6})is equivalent to Riccati equation
\begin{equation}\label{EQ3.7}
  \frac{d^2y}{dt^2}+\frac{\gamma^2}{4}
  \left(\frac{\bar{\rho}}{\rho_c}-1\right)y=0
\end{equation}
where $y=r\exp(-\gamma t/2)$ and
\begin{equation}\label{EQ3.8}
  \rho_c = \frac{3\gamma^2}{16\pi G},
\end{equation}
In WKB-approximation solution of Eq.(\ref{EQ3.7}) is
\begin{equation}\label{EQ3.8a}
  r \simeq \left\{
             \begin{array}{ccc}
               r_1e^{2\gamma t}\cos^4\phi(t)
               & for & t \ll t_* \\
               r_2\exp[\gamma (t-t_*)] & for & t\gg t_*
            \end{array}
          \right.
\end{equation}
where $r_{1,2}$ are the constants and
$\phi(t)$ should be found from equation
\begin{equation}\label{EQ3.9}
  \cos^6(\phi)\frac{d\phi}{dt} \simeq \frac{\gamma}{2}\sqrt{\frac{\rho_{int}}{\rho_c}}
  e^{-3\gamma t},
\end{equation}
and $t_*$ is estimated as
\begin{equation}\label{EQ3.10}
  t_* \sim \frac{1}{6\gamma}\ln\frac{\rho_{int}}{\rho_c},
\end{equation}
where $\rho_{int}$ is initial density of matter. We see that for $\rho_{int}\gg\rho_c$ and $t\ll t_*$ our Universe should be preceded by set of the closed universes, which are collapsed at the times $\phi(t_n)=(2n+1)\pi/2$.
Number of these precursor universes is estimated as
\begin{equation}\label{EQ3.11}
  N \sim \sqrt{\frac{\rho_{int}}{\rho_c}},
\end{equation}
and could be large. In the opposite case $t>t_*$ (where $\bar{\rho}<\rho_c$) we have open universe with accelerated expansion.

Solution of Eq.(\ref{EQ3.13}) can be roughly estimated as
\begin{equation}\label{EQ3.14}
  H \sim \left\{
          \begin{array}{lcc}
           2\gamma\left(1-\sqrt{\frac{\rho_{int}}{\rho_c}}
          \;e^{-3\gamma t}\tan\phi/\cos^6\phi\right)
          & for & t \ll t_* \\
          \gamma & for & t\gg t_*
          \end{array}
         \right.
\end{equation}
Numerical solution of (\ref{EQ3.13}) for $t>t_*$ is shown in Fig.~2. We see that $0.74\gamma\leq H(t>t_*)\leq \gamma$, so $\gamma$ can be estimated as $\gamma\simeq 2\cdot10^{-18}s^{-1}$, while deceleration parameter
\begin{equation}\label{EQ3.15}
  q=-1-\frac{\dot{H}}{H^2}=-\frac{\gamma}{H} + \frac{\gamma^2}{4H^2}\,\frac{\bar{\rho}}{\rho_c}.
\end{equation}
is $-1.2<q<0$ for $t>t_*$ that reasonable agree with experimental data \cite{Decelr} $q\simeq-1.08\pm0.29$ and indicates accelerated expansion of the matter.

It should be noted, that in our case the MF-approximation is equivalent to the classic limit of the Heisenberg equations. Accuracy of the MF-approximation can be estimated as
\begin{equation*}
  o\left(\frac{\overline{\Delta \bm{x}^2}}{r^2},
   \frac{\overline{\Delta \bm{V}^2}}{V^2}\right),
\end{equation*}
by using relation $\Delta \bm{V}^2=H^2\Delta \bm{x}^2$ and uncertainty relation
\begin{equation}\label{EQ3.16}
  \overline{\Delta \bm{x}^2}\; \overline{\Delta \bm{V}^2}
  \sim \frac{\hbar^2}{4m^2w^2}
\end{equation}
we obtain
\begin{equation}\label{EQ3.16a}
  \frac{\overline{\Delta \bm{x}^2}}{r^2} =
  \frac{\overline{\Delta \bm{V}^2}}{V^2}
  \sim \frac{\hbar}{2mwHr^2}.
\end{equation}
We see that for $t>t_*$ the MF-approximation is reasonable, while for epoch of the precursor universes it is failed near the collapse points, where, however, even exact Heisenberg equations (9),(10) have only qualitative sense, because the gravitation field is not weak.

\section{Kepler problem and spiral galaxies} \label{ApKepler}

Consider star with mass $m$ which is rotated around massive nucleus with mass $M$. Classical limit of the S-Heisenberg equations for this problem is
\begin{eqnarray}\label{EQ4.1}
  \frac{dR}{dt} &=& V_{\|}, \\ \label{EQ4.1a}
  \frac{d}{dt}(wV_{\|}) &=& -w\left[\frac{GM}{R^2}
   -\frac{V^2_{\perp}}{R}\right],\\ \label{EQ4.1b}
  \frac{d}{dt}(wV_{\perp})&=& - \frac{wV_{\|}V_{\perp}}
  {R}, \\  \label{EQ4.1b1}
  \frac{d\varphi}{dt} &=& \frac{V_{\perp}}{R}, \\ \label{EQ4.1c}
  \frac{dw}{dt} &=&-\gamma w,
\end{eqnarray}
where $R$ is distance of star from the nucleus, $V_{\|}$ is radial velocity, $V_{\perp}$ is rotational velocity and $\varphi$ is rotational angle. The system Eqs.(\ref{EQ4.1})-(\ref{EQ4.1c}) has a first integral
\begin{equation}\label{EQ4.2}
  wRV_{\perp}=const.
\end{equation}
which reflects conservation of the "super-angular-momentum".
For large $t$ and $R$ asymptotic solution for the distance is
\begin{eqnarray}\label{EQ4.3}
  R &\sim& e^{\gamma t} = w^{-1}, \\ \label{EQ4.3a}
  \varphi &\sim& 1-e^{-\gamma t},
\end{eqnarray}
which means that a galaxy can formate long spiral arms. Besides, Eqs.(\ref{EQ4.2}) and (\ref{EQ4.3}) show that at large distance from a nucleus the rotation velocity $V_{\perp}$ is tended to a constant, which agree with the observation \cite{Roten}. Numerical solution of Eqs.(\ref{EQ4.1})-(\ref{EQ4.1c}) is shown in Fig.~3.



\end{document}